\documentclass[10pt,letterpaper]{article}
\usepackage{opex3}
\usepackage{graphicx}
\usepackage{cite}

\begin{document}

\title{Fast optical cooling of nanomechanical cantilever with the dynamical
Zeeman effect}

\author{Jian-Qi Zhang$^{1}$, Shuo Zhang$^{2}$, Jin-Hua Zou$^{1,3}$, Liang
Chen$^{1}$, Wen Yang$^{4^\star}$, Yong Li$^{4*}$, and Mang Feng$^{1\dag}$}

\address{$^1$ State Key Laboratory of Magnetic Resonance and Atomic and Molecular Physics,
Wuhan Institute of Physics and Mathematics, Chinese Academy of Sciences - Wuhan National
Laboratory for Optoelectronics, Wuhan 430071, China\\
$^2$ College of Science, National University of Defense Technology, Changsha
410073, China\\
$^3$ College of Physical Science and Technology, Yangtze University, Jingzhou, 434023, China\\
$^4$ Beijing Computational Science Research Center, Beijing 100084, China}

\email{$^{\star}$wenyang@csrc.ac.cn, $^{*}$liyong@csrc.ac.cn, $^{\dag}$
mangfeng@wipm.ac.cn} %% email address is required



\begin{abstract} We propose an efficient optical electromagnetically induced transparency
(EIT) cooling scheme for a cantilever with a nitrogen-vacancy center
attached in a non-uniform magnetic field using dynamical Zeeman effect. In
our scheme, the Zeeman effect combined with the quantum interference effect
enhances the desired cooling transition and suppresses the undesired heating
transitions. As a result, the cantilever can be cooled down to nearly the
vibrational ground state under realistic experimental conditions within a
short time. This efficient optical EIT cooling scheme can be reduced to the
typical EIT cooling scheme under special conditions.\end{abstract}

 % REPLACE WITH CORRECT OCIS CODES FOR YOUR ARTICLE

%%%%%%%%%%%%%%%%%%%%%%% References %%%%%%%%%%%%%%%%%%%%%%%%%

%%%%%%%%%%%%%%%%%%%%%%%%%%  body  %%%%%%%%%%%%%%%%%%%%%%%%%%
\section{Introduction}

Micro- and nano-mechanical resonators (MRs) \cite%
{V.B.Braginsky,prl-97-237201} exhibit both classical and quantum properties
\cite{Physics-2-40}, which are relevant in fundamental physics and have
various applications, such as entanglement between mesoscopic objects \cite%
{pra-84-024301,prl-101-200503}, ultra-sensitive measurements \cite%
{Physics-2-40,prl-108-120801,pra-86-053806}, quantum information processing
\cite{prl-105-220501}, and biological sensing \cite{Nat.nanotech-3-501}.

Different from the classical properties of the MRs at room temperature, the
quantum properties of the MRs are presented only at sufficiently low
temperatures where the thermal fluctuations are suppressed. To suppress the
thermal fluctuations of the MRs and study their quantum properties, various
ground-state cooling schemes have been proposed, such as sideband cooling
\cite%
{prl-92-075507,Nature-475-359,PRB-78-134301,Nature-432-200,prl-99-093901,prl-99-093902,prb-76-205302}%
, cooling with static electrical interaction without any auxiliary qubit or
photonic systems \cite{JPCM-25-142201}, hot thermal light cooling \cite%
{prl-108-120602}, time-dependent control cooling \cite%
{JPCM-25-142201,pra-83-043804}, cooling method based on quadratic coupling
\cite{pra-85-025804}, measurement-based cooling \cite{PRB-84-094502}, and
dynamic dissipative cooling \cite{Dissipative-Cooling-PRL2013}.
Experimentally, efficient cooling of a MR with high frequency $\omega
_{m}\simeq 2\pi \times 6$ GHz down to the ground state (with an average
phonon number $\langle n\rangle \simeq 0.07$) was achieved with a direct
refrigerator of 25 mK \cite{Nature-464-697}. For most relevant MRs with
lower frequencies, only a few efficient cooling experiments have been
achieved with the final $\langle n\rangle <1$ \cite%
{Nature-475-359,Nature-478-89} via the sideband cooling scheme.

By contrast, in the cooling of vibrational motion of trapped ions instead of
MRs, there are some more efficient optical cooling schemes besides the
sideband cooling to achieve the ground-state cooling with the final average
phonon number much less than $\mathrm{1}$, such as the electromagnetically
induced transparency (EIT) cooling of vibrational motion of trapped ions
\cite{prl-85-5547,prl-85-4458} through the suppression of undesired
transitions \cite{prl-85-5547,prl-85-4458,njp-9-279}, and the Stark shift
cooling based on the Stark shift gate \cite{njp-9-279}. Those schemes remind
us of using the similar cooling schemes to cool the motion of massive MRs.

However, neither the EIT cooling \cite{prl-85-5547,prl-85-4458} nor the
Stark shift cooling \cite{njp-9-279} schemes in trapped ion systems can be
directly applied to the MR, since the direct optical Lamb-Dicke parameter
for the MR is too small to couple the vibrational motion of the MR with a
qubit for the small amplitude of the zero point fluctuation. Recently, a
\textit{microwave}-based EIT cooling scheme \cite{prl-103-227203} was
proposed in a MR electromechanical system, in which large effective
Lamb-Dick parameter can be achieved by controlling the applied magnetic
field on a flux qubit. An attractive alternative is to replace the flux
qubit with a nitrogen-vacancy (NV) center, which has important advantages
including operating temperatures from 4 to 300 K, stable fluorescence even
in small nanodiamonds, long spin lifetimes, optical initialization and
readout, biological compatibility, as well as available quantum memory that
can be encoded in proximal nuclear spins \cite{natnano}. The first MR
cooling scheme based on the NV center was proposed by Rabl \textit{et al.}
in Ref. \cite{prb-79-041302}, where it was shown that the
magnetic-tip-attached MR could be strongly coupled to the NV center through
a strong magnetic field gradient (MFG). The Lamb-Dicke parameter increases
with increasing MFG, making it possible to cool the MR with a small
frequency. This scheme is capable of producing an arbitrary quantum
superposition of the cantilever states \cite{prb-79-041302} and the
essential ingredient of this scheme, the strong MFG coupling, has already
been demonstrated in the experiment \cite{natphys-7-879}. Nevertheless, this
cooling scheme is based on \textit{microwave} and is efficient in the
\textit{resolved} sideband regime. It is desirable to develop an \textit{%
optical} cooling scheme that remains efficient in the \textit{non-resolved}
sideband regime, is faster and more efficient.

To this end, we present in this work an optical EIT cooling scheme for a
cantilever with a NV center attached, which can rapidly cool the cantilever
resonator down to nearly the vibrational ground state in the non-resolved
sideband regime. Our scheme consists of two essential ingredients which are
similar to Ref. \cite{prb-79-041302}. The first ingredient is the negatively
charged NV center attached at the end of a cantilever. The NV center serves
as a $\mathrm{\Lambda }$-type three-level system, exhibiting quantum
interference when two applied lasers are tuned to the two-photon resonance.
The second ingredient is the strong MFG, which can couple the internal
states of the NV center to the vibrational motion of the cantilever. Then,
with the external light fields applied, off-resonant and undesired carrier
transitions are suppressed, while the desired cooling transitions are
enhanced like in the typical EIT coolings \cite{prl-85-5547,prl-85-4458}. As
a result, our scheme can cool the vibrational motion of the cantilever close
to its ground state. The main components needed in our scheme are
experimentally available by using the current laboratory technology, e.g.,
the manipulation of a NV center with optical lights \cite{nature-466-730,
nature-478-497} and the coupling of a NV center with the nanowire \cite%
{natphys-7-879}.

Our cooling scheme differs significantly from the previous ones \cite{prb-79-041302,rmp-77-633,njp-9-279,prl-104-043003}. First, our
scheme uses optical lights to achieve an effective EIT cooling. The carrier
transition that dominates the heating process \cite{prl-103-227203} is
completely suppressed by the quantum interference effect, so it is efficient
in the non-resolved sideband regime, with the maximal cooling rate close to
the MFG coupling strength. By contrast, the previous NV center based
scheme \cite{prb-79-041302} uses microwaves and is efficient in the resolved
sideband regime, with the maximal cooling rate much smaller than the MFG
coupling strength. Second, our effective EIT cooling scheme uses the MFG
mechanical coupling of the resonator to the ground states of the NV center.
By contrast, the mechanical coupling in the typical atomic EIT cooling
scheme \cite{rmp-77-633} involves both the ground state and the excited
state of the atomic internal states. Third, compared with the Stark shift
cooling \cite{njp-9-279,prl-104-043003}, our scheme has a simpler
configuration, i.e., it does not require the light/microwave driven
transition between the two ground states.

The rest of the paper is organized as follows. In Sec. 2, we describe the
model, discuss the cooling process, and connect our Hamiltonian to the
typical EIT Hamiltonian by a canonical transformation. In Sec. 3, we derive
the analytical formula for the cooling and heating rates and calculate the
final average phonon number. In Sec. 4, we check the robustness and efficiency of
our scheme by comparing it to a fully numerical simulation. Finally, a brief
conclusion is given.

\section{Model and Cooling process}

The system under consideration is a negatively charged NV center attached at
the end of a nano-mechanical cantilever in a spatially non-uniform magnetic
field. The NV center consists of a substitutional nitrogen atom and a
neighboring carbon vacancy. The electronic ground state of the NV center is
an $S=1$ spin triplet with a zero field splitting of $\mathrm{2\pi \times
2.87}$ $\mathrm{GHz}$\ between the $m_{s}=0$ sublevel $\left\vert
0\right\rangle $ and the $m_{s}=\pm 1$ sublevels $\left\vert \pm
1\right\rangle $ due to spin-spin interactions, where $m_{s}$ is the
projection of the total electron spin $S=1$ along the $z$ (N-V) axis.
According to the group theoretical analysis \cite%
{njp-13-025025,nature-304-74}, we assume that a $\sigma ^{-}$- ($\sigma ^{+}$%
-) polarized laser with the Rabi frequency $\Omega _{-}$ ($\Omega _{+}$) and
light frequency $\omega _{-}$ ($\omega _{+}$ ) is applied to selectively
couple the ground state $\left\vert +1\right\rangle $ ($\left\vert
-1\right\rangle $) to the excited state $\left\vert A_{2}\right\rangle $
\cite{njp-13-025025,nature-304-74,pra-83-054306} as sketched in Fig. \ref%
{NVcooling}. Thus the three states $\left\vert \pm 1\right\rangle $ and $%
\left\vert A_{2}\right\rangle $ form a $\Lambda $-type three-level system,
which exhibits quantum interference when the frequencies of the two lasers
are tuned to two-photon resonance. In addition to decaying to the ground
states $|\pm 1\rangle $ by spontaneous emission, the excited state $%
|A_{2}\rangle $ also has a small probability to decay non-radiatively to the
metastable state $|^{1}A_{1}\rangle $ and then to the ground state $%
|0\rangle $. To prevent the leakage out of the $\Lambda $ system, we apply a
recycling laser to excite $|0\rangle $ to another excited state $%
|E_{y}\rangle $, which then decays back into the states $|\pm 1\rangle $. In Appendix A, we show that the effect of these recycling
transitions amounts to a small renormalization of the decay rate of $%
\left\vert A_{2}\right\rangle \rightarrow \left\vert +1\right\rangle $ ($%
\left\vert A_{2}\right\rangle \rightarrow \left\vert -1\right\rangle $) from
$\gamma _{+1}$ ($\gamma _{-1}$) to $\gamma _{+}$ $(\gamma _{-}$).

In the presence of a non-uniform magnetic field $B(\mathbf{r})$ along the
N-V symmetry axis of the NV center (defined as the $z$ axis), the Zeeman
effect $g_{e}\mu _{B}B(\mathbf{\hat{r}})(\left\vert +1\right\rangle
\left\langle +1\right\vert -\left\vert -1\right\rangle \left\langle
-1\right\vert )$ couples the NV center to the vibration of the NV center or
equivalently the vibration of the cantilever. Here, $\mathbf{\hat{r}}\equiv (%
\hat{x},\hat{y},\hat{z})$ is the position operator of the NV center attached
to the tip of the cantilever and $B(\mathbf{\hat{r}})$ is the magnetic field
on the NV center. As suggested in Refs. \cite{prl-104-043003, prl-87-257904}%
, the Stark shift can be caused by the Zeeman effect, while there is no
direct transition between the two ground states driven by an external light
field in our work. For this reason, our scheme differs from the Stark shift
cooling, but instead is an effective EIT cooling, i.e., it is connected to
the typical EIT cooling Hamiltonians \cite{prl-85-5547,prl-85-4458} by a
canonical transformation (to be shown below). Note that the excited state $%
\left\vert A_{2}\right\rangle $ is an equal mixture of $m_{s}=+1$ and $%
m_{s}=-1$ components, so its Zeeman effect is dominated by a second-order
process mediated by another excited state $\left\vert A_{1}\right\rangle $.
The large gap $\Delta _{A_{1}-A_{2}}\simeq $ $\mathrm{2\pi \times }$2.5 GHz
between $\left\vert A_{2}\right\rangle $ and $\left\vert A_{1}\right\rangle $
makes this second-order contribution negligible for a weak magnetic field $%
\left\vert B(\mathbf{\hat{r}})\right\vert \ll \hbar \Delta
_{A_{1}-A_{2}}/(g_{e}\mu _{B})\sim 10^{3}\ \mathrm{G}$. Taking the direction
of the cantilever vibration as the $x$ axis and the equilibrium position of
the cantilever tip (or equivalently the NV center) as the origin $\mathbf{r}%
=0$, the magnetic field $B(\mathbf{\hat{r}})=B(\hat{x},0,0)\equiv B(\hat{x})$
can be expanded as $B(\hat{x})\approx B(0)+B^{\prime }(0)\hat{x}$. The
zeroth-order term $g_{e}\mu _{B}B(0)(\left\vert +1\right\rangle \left\langle
+1\right\vert -\left\vert -1\right\rangle \left\langle -1\right\vert )$ is a
constant Zeeman splitting that lifts the degeneracy of $\left\vert \pm
1\right\rangle $. The first-order term $g_{e}\mu _{B}B^{\prime
}(0)(\left\vert +1\right\rangle \left\langle +1\right\vert -\left\vert
-1\right\rangle \left\langle -1\right\vert )\hat{x}$ induces the coupling
between the internal states of the NV center and the vibrational motion of
the cantilever through the strong MFG \cite{prb-79-041302,natphys-7-879},
where $\hat{x}=x_{0}(b+b^{\dagger })$ with $x_{0}=\sqrt{\hbar /(2M\omega
_{m})}$ the amplitude of the zero-point fluctuation of the cantilever with
mass $M$ and frequency $\omega _{m}$, and $b$ and $b^{\dagger }$ the
corresponding annihilation and creation operators, respectively.

Without loss of generality, we take the Rabi frequencies $\Omega _{\pm
}=\Omega _{0}$ as real numbers. Then, the total Hamiltonian for the coupled
system reads ($\hbar =1$)
\begin{equation}
\begin{array}{ccl}
H & = & \omega _{m}b^{\dagger }b+\omega _{A}\left\vert A_{2}\right\rangle
\left\langle A_{2}\right\vert +g_{e}\mu _{B}B(0)(\left\vert +1\right\rangle
\left\langle +1\right\vert -\left\vert -1\right\rangle \left\langle
-1\right\vert ) \\
& + & \frac{1}{2}\Omega _{0}(\left\vert A_{2}\right\rangle \left\langle
+1\right\vert e^{-i\omega _{+}t}+\left\vert A_{2}\right\rangle \left\langle
-1\right\vert e^{-i\omega _{-}t}+h.c.) \\
& + & \lambda (\left\vert +1\right\rangle \left\langle +1\right\vert
-\left\vert -1\right\rangle \left\langle -1\right\vert )(b^{\dagger }+b).%
\end{array}
\label{C0}
\end{equation}%
The first line describes the free evolution of the cantilever and the NV
center with $\omega _{A}$ being the energy of state $|A_{2}\rangle $. The
second line describes the selective excitation of the internal states of the
NV center by two laser fields. The last line describes the MFG induced
coupling between the internal states of the NV center and the vibration of
the cantilever. The coupling strength $\lambda =$ $g_{e}\mu _{B}B^{\prime
}(0)x_{0}$ is equal to the change of the Zeeman splitting over a zero-point
fluctuation $x_{0}$. The physical picture for this interaction is that the
first-order term $g_{e}\mu _{B}B^{\prime }(0)\hat{x}$, depending on the
position of the NV center $\hat{x}$, induces an additional Zeeman splitting
between the ground states $\left\vert +1\right\rangle $ and $\left\vert
-1\right\rangle $. For a silicon cantilever with $(\mathrm{%
length,width,thickness})\simeq (25,0.1,0.1)\ \mathrm{\mu m}$, the cantilever
vibrational frequency $\omega _{m}=2\pi \times 1$ \textrm{MHz}, the mass $%
M=1.22\times 10^{-14}$ \textrm{kg}, and the zero-point amplitude $%
x_{0}\simeq 1.6\times 10^{-13}\ \mathrm{m}$. Thus the coupling strength can
reach $\lambda \simeq 2\pi \times 0.115\ \mathrm{MHz}$ \cite%
{prb-79-041302,nphys-6-602} for an achievable MFG value $B^{\prime }(0)\sim
2.4\times 10^{7}\ \mathrm{T/m}$. Consequently, the magnetic Lamb-Dicke
parameter $\eta \equiv \lambda /\omega _{m}=g_{e}\mu _{B}B^{\prime
}(0)x_{0}/\omega _{m}$ can be as large as $\eta \sim 0.115$ by using a
strong MFG even for a massive MR with a small zero-point fluctuation $x_{0}$%
. This is in sharp contrast to the optical Lamb-Dicke parameter $\eta _{%
\mathrm{ion}}=2\pi x_{0}/\lambda _{\mathrm{light}}$ for the trapped ion,
which scales as $x_{0}$ over the optical wavelength $\lambda _{\mathrm{light}%
}$. The MFG coupling of the NV center and its instantaneous influence on the
NV center has been observed experimentally \cite{natphys-7-879}.

\begin{figure}[tbp]
\centering\includegraphics[width=5cm]{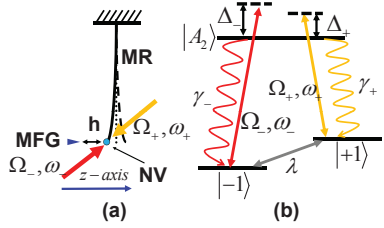}
\caption{(a) Schematic illustration of the optical scheme of NV
center-assisted scheme of the effective optical EIT cooling. (b) Three
internal energy levels of the NV center are coupled by two lasers satisfying
the two-photon resonance $\Delta _{+}=\Delta _{-}$. The cantilever vibration
is coupled to the NV center by a strong MFG. $\protect\gamma _{+}$ ($\protect%
\gamma _{-}$) is the decay from the excited state $|A_{2}\rangle $ to the
ground state $|-1\rangle $ ($|+1\rangle $), which includes the electron
leaking from and pumped into the nearly closed three-level for $%
|A_{2}\rangle $ and $|\pm 1\rangle $ (see Appendix A for details).}
\label{NVcooling}
\end{figure}

In the rotating frame defined by $|\psi ^{\mathrm{rot}}(t)\rangle
=e^{-iRt}|\psi (t)\rangle $ and $H^{\mathrm{rot}}=e^{-iRt}H(t)e^{iRt}+R$
with $R\equiv \omega _{+}\left\vert +1\right\rangle \left\langle
+1\right\vert +\omega _{-}\left\vert -1\right\rangle \left\langle
-1\right\vert $ \cite{Gardiner}, the Hamiltonian is time-independent:
\begin{equation}
\begin{array}{ccl}
H^{\mathrm{rot}} & = & \omega _{m}b^{\dagger }b-\Delta \left\vert
A_{2}\right\rangle \left\langle A_{2}\right\vert +\frac{1}{2}\Omega
_{0}(\left\vert A_{2}\right\rangle \left\langle +1\right\vert +\left\vert
A_{2}\right\rangle \left\langle -1\right\vert +h.c.) \\
& + & \lambda (\left\vert +1\right\rangle \left\langle +1\right\vert
-\left\vert -1\right\rangle \left\langle -1\right\vert )(b^{\dagger }+b),%
\end{array}
\label{C01}
\end{equation}%
where the detunings $\Delta _{\pm }\equiv \omega _{\pm }-[\omega _{A}\mp
g_{e}\mu _{B}B(0)]$ of the $\sigma ^{\pm }$ lasers have been chosen to
satisfy the two-photon resonance condition $\Delta _{+}=\Delta _{-}\equiv
\Delta $.

Then we make a canonical transformation $H^{\mathrm{rot}}\rightarrow H_{%
\mathrm{e}}\equiv e^{-iS}H^{\mathrm{rot}}e^{iS}$ with $S=-i\eta (\left\vert
+1\right\rangle \left\langle +1\right\vert -|-1\rangle \left\langle
-1\right\vert )\left( b-b^{\dagger }\right) $. Up to the first-order of the
small Lamb-Dicke parameter $\left\vert \eta \right\vert \ll 1$, the
Hamiltonian $H_{\mathrm{e}}$ takes the similar form as the one in the
typical EIT cooling scheme \cite{prl-85-5547,prl-85-4458} as $H_{\mathrm{e}%
}(t)=H_{\mathrm{0}}+V$, in which
\begin{equation}
H_{\mathrm{0}}\equiv \omega _{m}b^{\dagger }b-\Delta \left\vert
A_{2}\right\rangle \left\langle A_{2}\right\vert +\frac{\sqrt{2}}{2}\Omega
_{0}(\left\vert A_{2}\right\rangle \left\langle b\right\vert +h.c.),
\label{C1}
\end{equation}%
describes the motion of the NV center driven by the two lasers and
\begin{equation}
V=\eta \left( b-b^{\dagger }\right) (\frac{\Omega _{0}}{\sqrt{2}}\left\vert
A_{2}\right\rangle \left\langle d\right\vert -h.c.),  \label{C2}
\end{equation}%
is the coupling between the NV center and the cantilever. It is well known
that the Hamiltonian of the NV center (in the absence of the MR) can be
diagonalized as $H_{\mathrm{NV}}=E_{+}\left\vert +\right\rangle \left\langle
+\right\vert +E_{-}\left\vert -\right\rangle \left\langle -\right\vert $ by
the dark state $\left\vert d\right\rangle $ (with zero eigenenergy) and the
two dressed states $\left\vert +\right\rangle =\cos \phi \left\vert
A_{2}\right\rangle +\sin \phi \left\vert b\right\rangle $ and $\left\vert
-\right\rangle =\sin \phi \left\vert A_{2}\right\rangle -\cos \phi
\left\vert b\right\rangle $, where $\phi =(1/2)\cos ^{-1}(-\Delta /\sqrt{%
2\Omega _{0}^{2}+\Delta ^{2}})$\ and the eigenenergies $E_{\pm }\equiv
(-\Delta \pm \sqrt{2\Omega _{0}^{2}+\Delta ^{2}})/2$\ \cite{pra-72-043823}.
The linewidths of the two dressed states $\left\vert +\right\rangle $ and $%
\left\vert -\right\rangle $ are $\Gamma \cos ^{2}\phi $ and $\Gamma \sin
^{2}\phi $, respectively, which means that all the decay result from the
decay $\Gamma =\gamma _{+}+\gamma _{-}$ from the excited state $\left\vert
A_{2}\right\rangle $. In this situation, the state with the lowest energy is
$\left\vert -\right\rangle $, whose dominant component is $\left\vert
b\right\rangle $\ (for $\Delta <0$) or $\left\vert A_{2}\right\rangle $\
(for $\Delta >0$).

\begin{figure}[tbp]
\centering\includegraphics[width=6cm]{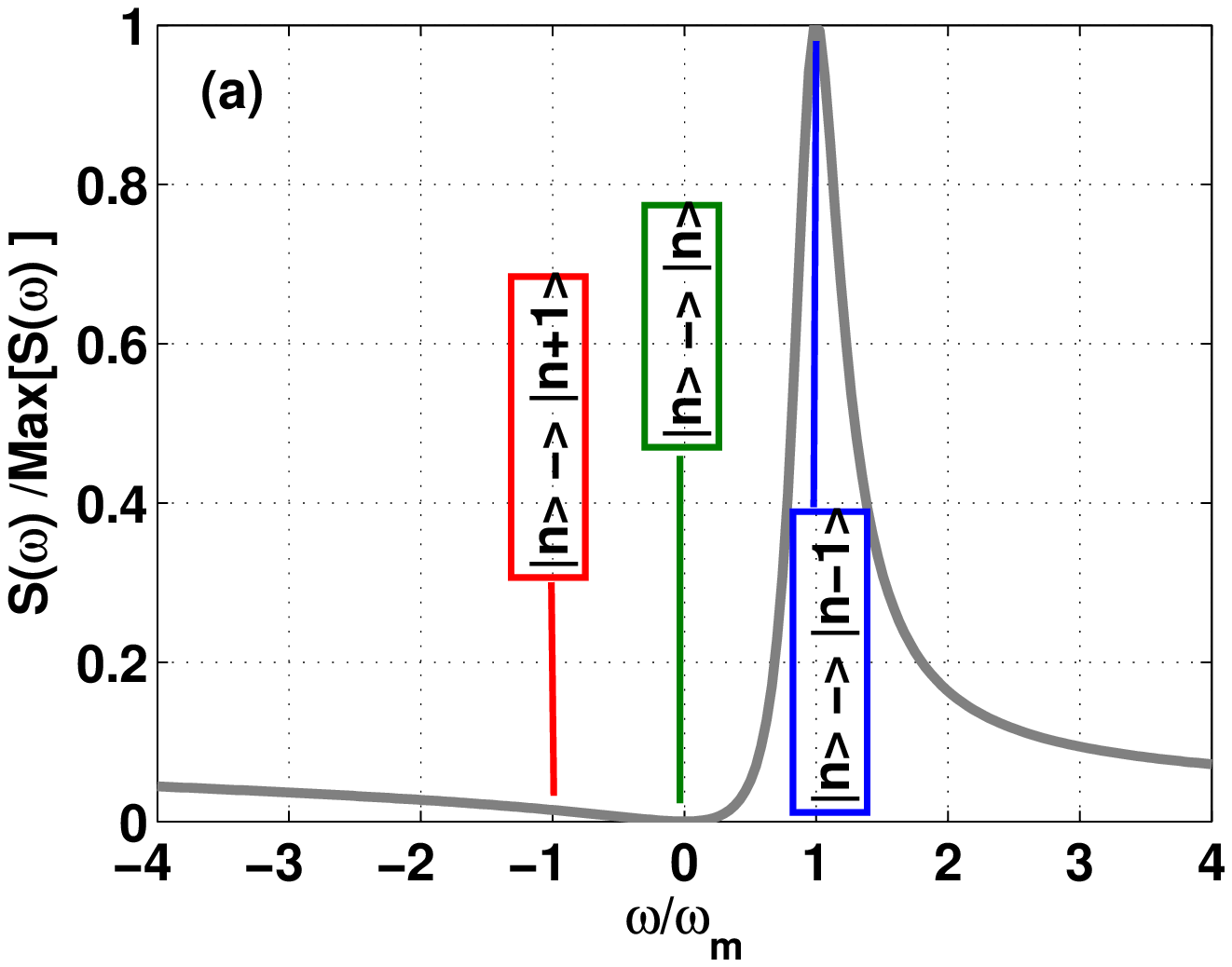} \centering%
\includegraphics[width=5cm]{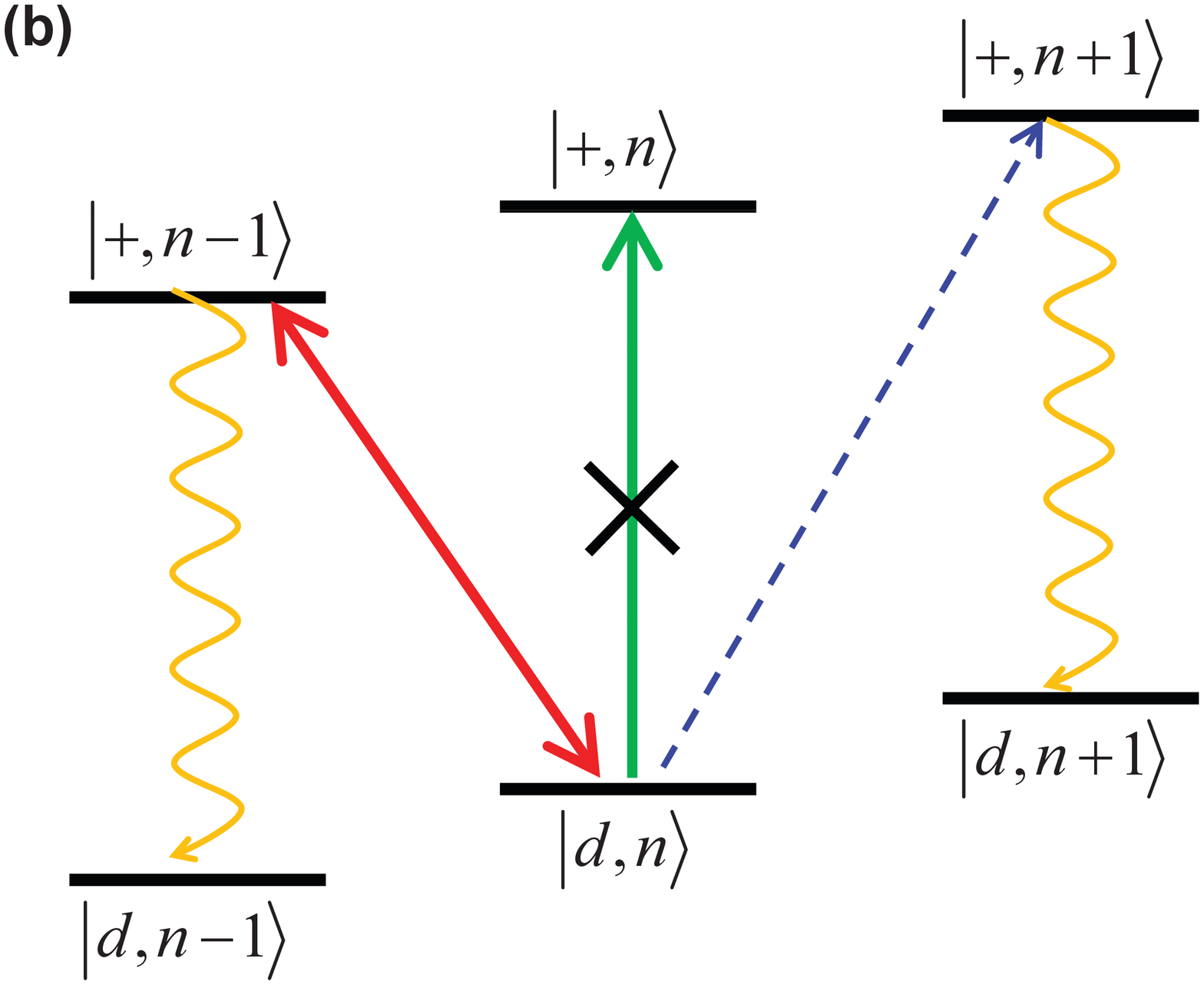}
\caption{(a) Absorption spectrum versus the frequency in our scheme. It is
demonstrated that there is a dark dip for two-photon resonance in the
absorption rate, and the quantum interference can be used to suppress the
heating transitions $|n\rangle\rightarrow|n\rangle$ and $|n\rangle%
\rightarrow|n+1\rangle$ and enhance the cooling transition $%
|n\rangle\rightarrow|n-1\rangle$. The parameters are taken as $\protect\omega%
_{m}=2\protect\pi\times1$ MHz, $\Gamma=15 \protect\omega_m$ , $\Omega_0=8%
\protect\omega_m$, and $\Delta=31\protect\omega_m$; (b) The cooling cycle of
our scheme. The action of the external light fields (with the Rabi
frequencies $\Omega _{0}$) and the Zeeman effect (with the coupling $\protect%
\lambda $) can only create the red sideband transition $\left\vert
d,n\right\rangle \rightarrow \left\vert +,n-1\right\rangle $ since the
carrier transition between $\left\vert d,n\right\rangle $ and $\left\vert
+,n\right\rangle $ and a blue sideband transition $\left\vert
d,n\right\rangle $ and $\left\vert +,n+1\right\rangle $ are suppressed when
two applied lasers are tuned to two-photon resonance. Then, if the decay is
from $\left\vert +,n-1\right\rangle $ to $\left\vert d,n-1\right\rangle $,
one phonon has been lost compared with the initial state, whereas if the
transition is $\left\vert +,n-1\right\rangle\rightarrow \left\vert
d,n\right\rangle $, the cycle will be repeated. }
\label{cool}
\end{figure}

As sketched in Fig. \ref{cool}, the cooling process in our scheme can be
understood as a typical EIT cooling \cite{prl-85-5547,prl-85-4458}. From the
initial state $\left\vert d,n\right\rangle $, the action of the external
light fields (with the Rabi frequencies $\Omega _{0}$) and the Zeeman effect
(with the coupling $\lambda $) can only create the red sideband transition $%
\left\vert d,n\right\rangle \rightarrow \left\vert +,n-1\right\rangle $
since both the carrier transition between $\left\vert d,n\right\rangle $ and
$\left\vert +,n\right\rangle $ and the blue sideband transition between $%
\left\vert d,n\right\rangle $ and $\left\vert +,n+1\right\rangle $ are
suppressed when two applied lasers are tuned to two-photon resonance [see
Fig. \ref{cool}(a)]. Then, if the decay is from $\left\vert
+,n-1\right\rangle $ to $\left\vert d,n-1\right\rangle $, one phonon has
been lost compared with the initial state, whereas if the transition is $%
\left\vert +,n-1\right\rangle \rightarrow \left\vert d,n\right\rangle $, the
cycle will be repeated. Therefore, the mean phonon number decreases as the
phonons are rapidly dissipated into the thermal bath.

According to the description above, the efficient EIT cooling in our work is
based on the Zeeman effect caused by a MFG \cite{prl-104-043003} in the
ground states, different from the typical EIT cooling based on a constant
magnetic field with the transitions containing the excited state \cite%
{prl-103-227203}. In addition, compared with Ref. \cite{prl-104-043003}, our
model is a reduced model since there is no external light field driving the
transition between the two ground states in the Hamiltonian (\ref{C0}), and
the optical Lamb-Dicke parameters are too weak to generate the mechanical
effect of the light on the cantilever.

\section{The analytical result for the final mean phonon number}

Using the perturbation theory and the non-equilibrium
fluctuation-dissipation relation based on Eqs. (\ref{C1}) and (\ref{C2}), we
may derive the heating (cooling) coefficient $A_{+}$ ($A_{-}$) as (see
Appendix B for details)
\begin{equation}
A_{\pm }=2\Gamma \eta ^{2}\Omega _{0}^{2}\frac{\omega _{m}^{2}}{\Gamma
^{2}\omega _{m}^{2}+4(\Omega _{0}^{2}/2\pm \Delta \omega _{m}-\omega
_{m}^{2})^{2}},  \label{C21}
\end{equation}%
which is the same as the heating (cooling) coefficient given in Refs. \cite%
{prl-85-4458, prl-103-227203}, but different from the heating (cooling)
coefficient in Ref. \cite{njp-9-279,prl-104-043003} since there is no
external light or microwave field driving the transition between two ground
states directly. The rate $A_{+}$ corresponds to the heating transition $%
|d,n\rangle \rightarrow |\pm ,n+1\rangle $, which is resonant when $E_{\pm
}+\omega _{m}=0$. The rate $A_{-}$ corresponds to the transition $%
|d,n\rangle \rightarrow |\pm ,n-1\rangle $, which is resonant when $E_{\pm
}-\omega _{m}=0$.

Compared with trapped ions, the cantilever is more sensitive to the environmental
noise. It is reasonable to consider the decay caused by the thermal bath.
Then the following rate equation for the phonon occupation probability $P(n)$
on each Fock state $\left\vert n\right\rangle $ can be constructed:
\begin{equation}
\begin{array}{lll}
\frac{d}{dt}P(n) & = & [A_{-}+(N(\omega _{m})+1)\gamma
_{m}][(n+1)P(n+1)-nP(n)] \\
& + & [A_{+}+N(\omega _{m})\gamma _{m}][nP(n-1)-(n+1)P(n)],%
\end{array}
\label{C202}
\end{equation}%
where $\gamma _{m}=\omega _{m}/Q$ is the decay of the cantilever with $Q$
being the quality of the cantilever, and $N(\omega _{m})=[\exp (\omega
_{m}/k_{B}T)-1]^{-1}$ is the thermal occupation of the cantilever \cite%
{prl-103-227203}. The terms containing $N(\omega _{m})$ in Eq. (\ref{C202})
represent the additional cooling and heating coefficients from the thermal
bath. Then, the corresponding equation for the average phonon number $%
\left\langle n\right\rangle $ is \cite{prl-103-227203}
\begin{equation}
\frac{d}{dt}\left\langle n\right\rangle =-(W+\gamma _{m})\left\langle
n\right\rangle +A_{+}+N(\omega _{m})\gamma _{m},  \label{C9}
\end{equation}%
with $W=A_{-}-A_{+}$ being the net cooling rate induced by the NV center
\cite{prl-85-4458}.

The solution to the time-dependent average phonon number is
\begin{equation}
\left\langle n(t)\right\rangle =\left\langle n\right\rangle
_{ss}+e^{-(W+\gamma _{m})t}[N(\omega _{m})-\left\langle n\right\rangle
_{ss}],  \label{C90}
\end{equation}%
where
\begin{equation}
\begin{array}{lll}
\left\langle n\right\rangle _{ss} & = & [A_{+}+N(\omega _{m})\gamma
_{m}]/(W+\gamma _{m}) \\
& \simeq  & A_{+}/W+N(\omega _{m})\gamma _{m}/W,%
\end{array}
\label{C10}
\end{equation}%
is the final average phonon number for Eq. (\ref{C9}) under the condition of
$W\gg \gamma _{m}$.

\begin{figure}[tbph]
\centering\includegraphics[width=6cm]{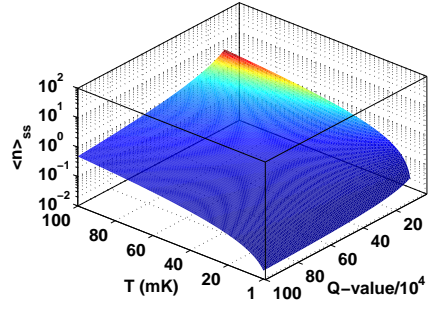}
\caption{(Color online) The final phonon number $\mathrm{log}\left\langle
n\right\rangle _{ss}$ versus $Q$-value and environmental temperature T (mk),
where we have $\protect\omega _{m}=2\protect\pi \times 1$ MHz, $\Omega _{0}=8%
\protect\omega _{m}$, $\Delta =31\protect\omega _{m}$, $\Gamma =15\protect%
\omega _{m}$, and $\protect\eta =0.115$ \protect\cite%
{prb-79-041302,natphys-7-879,nature-466-730,njp-13-025025,pra-83-054306,arxiv1305.1701}%
.}
\label{simu}
\end{figure}

The physical pictures for the two equations above are very clear. Eq. (\ref%
{C90}) shows that the phonon number decreases monotonically to its steady
state value $\left\langle n\right\rangle _{ss}$, indicating that coupling to
the optically pumped NV center increases the dissipation of the cantilever.
If the initial average photon number $\left\langle n\right\rangle
>\left\langle n\right\rangle _{ss}$, then the cooling $(W+\gamma
_{m})\left\langle n\right\rangle $ dominates over the heating $%
A_{+}+N(\omega _{m})\gamma _{m}$ and the average phonon number will
decrease, until it becomes equal to $\left\langle n\right\rangle _{ss}$ and
the cooling process stops. Equation (\ref{C10}) shows that the steady-state
phonon number $\left\langle n\right\rangle _{ss}$ is determined by both the $%
Q$-value ($Q=\omega _{m}/\gamma _{m}$) of the cantilever and the
environmental temperature. The cantilever is easily heated when the
cantilever decay rate $\gamma _{m}$ or environmental temperature increases.
This can be further understood by checking Fig. \ref{simu}, where the final
mean photon number increases with the decrease of $Q$-value (increase of
environmental temperature) for a given environmental temperature ($Q$%
-value). The final average phonon number in Eq. (\ref{C10}) is different
from that of Ref. \cite{prb-79-041302} because of the different dissipation
dynamics. In our scheme, the dissipation is from the excited state $%
\left\vert A_{2}\right\rangle $ to the two ground states $\left\vert \pm
1\right\rangle $. It is different from that in Ref. \cite{prb-79-041302},
where the dissipations are from the two higher states $\left\vert \pm
1\right\rangle $ to the lower state $\left\vert 0\right\rangle $.

\begin{figure}[tbph]
\centering\includegraphics[width=7cm]{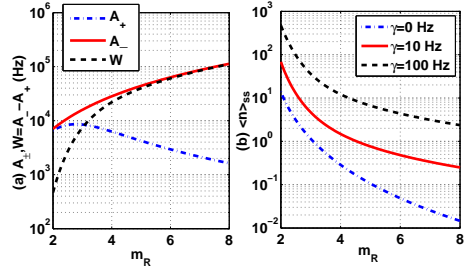}
\caption{(a) The cooling and heating coefficients $A_{\pm}$ and the cooling
rate $W$ versus the ratio $m_{R}=\Omega_{0}/\protect\omega_{m}$. Here the
blue dotted, red solid, and black dashed lines correspond to the heating
coefficient, cooling coefficient, and cooling rate, respectively. (b) The
final average phonon number $\left\langle n\right\rangle_{ss}$ versus the
ratio $m_{R}$. The parameters are taken from Refs. \protect\cite%
{prb-79-041302,natphys-7-879,nature-466-730,njp-13-025025,pra-83-054306,arxiv1305.1701}
as $\protect\omega_{m}=2\protect\pi\times 1$ MHz, $\Gamma=15 \protect\omega%
_m $, $T=20$ mK and $\protect\eta=0.115$.}
\label{steady}
\end{figure}

We define a ratio $m_{R}=\Omega _{0}/\omega _{m}$, which means the Rabi
frequency $\Omega _{0}$ increases with the increase of the ratio $m_{R}$. To
have the largest cooling coefficient $A_{-}$, we choose $\Delta
=(m_{R}^{2}-2)\omega _{m}/2$ which can ensure the transition $\left\vert
d,n\right\rangle \rightarrow \left\vert +,n-1\right\rangle $ is resonant.
Then we may rewrite the heating and cooling coefficients as
\begin{equation}
A_{+}=\eta ^{2}\frac{2m_{R}^{2}\omega _{m}^{2}\Gamma }{4(m_{R}^{2}-2)^{2}%
\omega _{m}^{2}+\Gamma ^{2}},~~~~~A_{-}=\eta ^{2}\frac{2m_{R}^{2}\omega
_{m}^{2}}{\Gamma },  \label{C11-2}
\end{equation}%
which imply a hump curve in the heating coefficient and a parabolic curve
for the cooling coefficient, as functions of $m_{R}$ [See Fig. \ref{steady}%
(a)]. With assistance of Eqs. (\ref{C1}) and (\ref{C2}), these phenomena can
be understood from the EIT cooling \cite{prl-85-4458} as follows. Although
the transition rates for both heating and cooling increase with the Rabi
frequencies, the linewidth of the dressed state $\left\vert +\right\rangle $
decreases with the increase of detuning, and the states $\left\vert
+,n-1\right\rangle $, $\left\vert +,n\right\rangle $ and $\left\vert
+,n+1\right\rangle $ can be distinguished more clearly. Therefore, the red
sideband transition between $\left\vert d,n\right\rangle $ and $\left\vert
+,n-1\right\rangle $ is enhanced due to the resonance caused by the Rabi
frequencies, and the carrier (blue sideband) transition between $\left\vert
d,n\right\rangle $ and $\left\vert +,n\right\rangle $ ($\left\vert
+,n+1\right\rangle $) is strongly suppressed by quantum interference. But
the transition between $\left\vert d,n\right\rangle $ and $\left\vert
+,n+1\right\rangle $ is enhanced first with the increase of Rabi
frequencies, and then suppressed due to the detuning caused by the Rabi
frequencies. As a result, the cooling coefficient is monotonically
increasing, but the heating coefficient presents a hump with the increase of
Rabi frequency.

Moreover, Fig. \ref{steady}(a) also shows that the cooling rate $%
W=A_{-}-A_{+}$ is close to the cooling coefficient $A_{-}$ when $m_{R}\geq 8$
because the cooling coefficient is over ten times larger than the heating
one in that case, e.g., the cooling coefficient $A_{-}=112.9 $ kHz and the
heating coefficient $A_{+}=1.6$\ kHz if $m_{R}=8$.

\begin{figure}[tbph]
\centering\includegraphics[width=6cm]{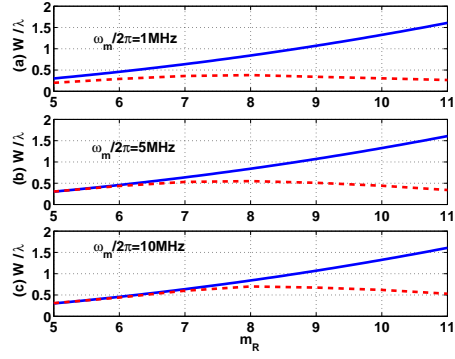}
\caption{The cooling $W$ in units of the magnetic coupling strength $\protect%
\lambda $ versus the ratio $m_{R}=\Omega _{0}/\protect\omega _{m}$. Here the
blue red solid and red dashed lines correspond to the analytic cooling rate $%
W=A_{-}-A_{+}$ for Eq.(\protect\ref{C11-2}) and the numerical cooling rate
from the master equation for Eq.(\protect\ref{C01}), respectively. And the
frequency of the cantilever $\protect\omega _{m}/2\protect\pi $ in (a), (b),
and (c) are 1\textrm{MHz}, 5\textrm{MHz}, and 10\textrm{MHz}. Here, $\Gamma
/2\protect\pi =15$\textrm{MHz}, $\protect\lambda /2\protect\pi =0.1$\textrm{%
MHz}. }
\label{result}
\end{figure}

%In addition, it is worth mentioning that, our scheme works in the
%non-resolved regime ($\Gamma >\omega _{m}$), and the largest numerical
%cooling rate $W\simeq 0.7\lambda $ [see Fig. \ref{result}(c)] can be much
%lager than the one $W\simeq 0.8\frac{\lambda ^{2}}{\Gamma }(\frac{\Gamma
%\Gamma _{0}}{\Omega _{0}^{2}})\ll 0.8\lambda $ in Ref. \cite{prb-79-041302}
%which works in the resolved regime ($\Gamma <\omega _{m}$). Here, $\Gamma
%_{0}$\ is the decay from the excited state $|E_{y}\rangle $\ to the ground
%state $|0\rangle $ with $\Gamma _{0}\simeq 100\omega _{m}$ \cite%
%{nature-478-497}. As a result, compared with Ref. \cite{prb-79-041302}, our
%scheme can realize a faster cooling.

It is worth mentioning that the adiabatic requirement $\Gamma \gg \lambda $
for deriving our analytical cooling rate formula Eq. (\ref{C21}) is always
satisfied, due to the very large NV center excited state decay rate $\Gamma
/(2\pi )=15$ MHz compared with the achievable value of $\lambda /(2\pi )=0.1$
MHz. It is interesting to notice that our analytical formula suggests that
the cooling rate $W$ can exceed the NV-cantilever coupling strength $\lambda
$ for sufficiently large optical Rabi frequency $\Omega _{0}$. We check
whether this is true by plotting in Fig. \ref{result} the cooling rate
(obtained from direct numerical simulation of the coupled evolution) versus
the Rabi frequency $\Omega _{0}$ for different cantilever frequencies $%
\omega _{m}$. Figure \ref{result} shows that the maximal cooling rate is
still limited by the coupling strength, e.g., $W\simeq 0.7\lambda $ ($%
0.55\lambda $, $0.38\lambda $) for the cantilever frequency $\omega
_{m}/2\pi =10\ $\textrm{MHz} (5\ \textrm{MHz}, 1\ \textrm{MHz}) and $\Omega
_{0}=8\omega _{m}$. Nevertheness, the maximal cooling rate in the
non-resolved sideband regime $\Gamma >\omega _{m}$ can be close to the
coupling strength, e.g., $W\simeq 0.7\lambda $ at $\omega _{m}/2\pi =$10\
\textrm{MHz} [see Fig. \ref{result} (c)]. This is to be contrasted with Ref.
\cite{prb-79-041302}. There, efficient cooling is achieved in the resolved
sideband regime and the cooling rate $W\simeq 0.8\lambda ^{2}/\Gamma
_{op}\ll 0.8\lambda $ due to the adiabatic requirement $\Gamma _{op}\gg
\lambda $, i.e., the effective damping rate $\Gamma _{op}$ of the NV center
should be much larger than the NV-cantilever coupling $\lambda $.

The corresponding final average phonon number as a function of the ratio $%
m_{R}$ is plotted in Fig. \ref{steady}(b) for different cantilever decay $%
\gamma _{m}$, which is given by
\begin{equation}
\left\langle n\right\rangle _{ss}=\frac{\Gamma ^{2}}{16[(m_{R}^{2}-2)/2%
\omega _{m}]^{2}}+N(\omega _{m})\frac{\gamma _{m}}{W}=\left( \frac{\Gamma }{%
4\Delta }\right) ^{2}+N(\omega _{m})\frac{\gamma _{m}}{W}.  \label{C12}
\end{equation}%
It means that the final average phonon number $\left\langle n\right\rangle
_{ss}$ decreases with the increase of the detuning $\Delta $ for the fixed
decay when the cooling coefficient $A_{-}$ takes its corresponding maximum
value. Furthermore, Fig. \ref{steady}(b) also shows that a lower final mean
phonon number can be obtained by decreasing the cantilever decay rate.
Besides, the equation for this cooling limit (\ref{C12}) is identical to the
one in EIT cooling \cite{prl-85-5547,prl-85-4458,prl-103-227203}.

\section{Simulation and Discussion}

We have demonstrated in the above sections the possibility of the efficient
optical EIT cooling for a cantilever attached with a NV center under the
strong gradient magnetic field and laser fields. The cantilever can be
cooled into its ground state by using the quantum interference and Zeeman
effect. In the following, we will give some simulations to discuss how the
scheme works in a realistic system.
\begin{figure}[tbph]
\centering\includegraphics[width=6cm]{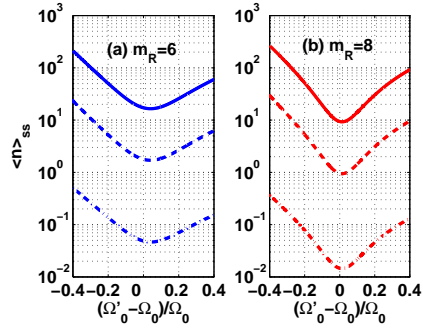} \centering
\caption{The final average phonon number $\langle n\rangle _{ss}$ as a
function of the variations around the optimal Rabi frequency $\Omega
_{0}=m_{R}\protect\omega _{m}$ with the detuning $\Delta =(m_{R}^{2}-2)%
\protect\omega _{m}/2$. The simulation is made by Eq.(\protect\ref{C10}),
where $\protect\omega _{m}=2\protect\pi \times 1$MHz, $\Gamma =15\protect%
\omega _{m}$, $T=20$mK, $\protect\eta =0.115$ \protect\cite%
{prb-79-041302,natphys-7-879,nature-466-730,njp-13-025025,pra-83-054306,arxiv1305.1701}%
. The dash line, dot-dash line and solid line correspond to $\protect\gamma %
_{m}=0$ Hz, $\protect\gamma _{m}=10$ Hz, and $\protect\gamma _{m}=100$ Hz,
respectively.}
\label{omegaton}
\end{figure}
To understand how the scheme works in a realistic system with the influence
of the Rabi frequency, we plot in Fig. \ref{omegaton} the final phonon
number by introducing fluctuation in the Rabi frequency $\Omega _{0}$. Here,
we choose the parameters from the experimentally achievable constants in
nano-mechanics \cite{prb-79-041302,natphys-7-879,njp-13-025025,pra-83-054306}
as follows: $\omega _{m}=\mathrm{2\pi \times 1}$ MHz, $\Gamma =\mathrm{%
15\omega }_{m}$, $T=\mathrm{20}$ mK and $\eta =0.115$, and $Q=10^{5}$. All
these parameters meet the approximation conditions mentioned above.

Figure \ref{omegaton} shows that the minimum value of the final average
phonon number deviates from the point of $\Omega_{0}^{\prime }=\Omega _{0}$.
This is due to the fact that the ideal cooling in our scheme depends on
maximum value of $A_{-}$, rather than the maximum value of $A_{-}/A_{+}$.
Moreover, Fig. \ref{omegaton} also shows that the final average phonon
number is more sensitive to the experimental error in the case of the larger
Rabi frequency since in this case the small deviation yields a large
detuning if $(\Omega _{0}^{\prime }-\Omega _{0})/\Omega _{0}$ is ascertained.

As a result, to cool the cantilever down to its vibrational ground state, we
must elaborately control the experimental imperfection in implementing our
scheme. In addition, since the cantilever is sensitive to the environment,
particularly for the small frequency cantilever, we have to remain the
cooling by keeping irradiation from the external light fields.

Moreover, it is worth to point out that the nuclear spin bath gives a
considerable influence on the final average phonon number and cooling time.
For example, under the parameters with $\langle n \rangle_{initial}=20$, $%
m_R=8$ and $\gamma =$\textrm{10}Hz as in Fig. \ref{omegaton}, the final
average phonon number is increased from $\left\langle n\right\rangle _{ss}=$%
0.955 to $\left\langle n\right\rangle _{ss}=$1.111 (10.576) when the nuclear
spin bath takes the random energy $\delta _{n}\leq 2\pi\times$ 0.1 MHz (0.5
MHz). The corresponding cooling time is increased from $t=0.547ms$ to $%
t=0.642ms (4.317ms)$, respectively. To decrease the impact of the nuclear
spin bath, we should use the dynamic nuclear polarization technology \cite%
{prl-102-057403} and the isotopic purification of NV center \cite{NL-12-2083}
to overcome this problem.

\section{Conclusion}

In summary, we have presented a protocol to cool the cantilever with a NV
center attached down to the vibrational ground state under the strong MFG
and laser fields. During the cooling process, the heating effects caused by
the carrier transition and the blue sideband transition can be suppressed by
quantum interference, and the cooling effects caused by the red sideband can
be enhanced by increasing Rabi frequencies. We have shown the possibility of
our scheme to cool the cantilever close to its ground state by controlling
experimental imperfection. This implies that our scheme is a good candidate
to realize the fast ground state cooling of cantilevers. In addition, we
have also proven that our efficient optical EIT cooling proposal can be
reduced to the EIT cooling one under some special conditions.

\section*{Appendix A}

\begin{figure}[tbph]
\centering\includegraphics[width=5cm]{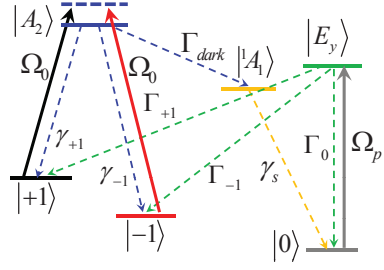} \centering
\caption{The pumping process and the decays of NV center. The transition
from state $|0\rangle$ and state $|E_{y}\rangle$ is driven by the pumping
light $\Omega_p$. Here $\protect\gamma_{\pm1}$ and $\Gamma_{dark}$ are the
direct decays from the excited state $|A_2\rangle$ to the ground states $%
|\pm1\rangle$ and the metastable state $|^{1}A_1\rangle$, respectively. $%
\Gamma_{\pm1}$ and $\Gamma_{0}$ are the direct decays from the excited state
$|E_y\rangle$ to the ground states $|\pm1\rangle$ and $|0\rangle$,
respectively, while $\Gamma_{op}^{\pm1}$ is the indirect decay from the
state $|0\rangle$ to the state $|\pm1\rangle$. $\protect\gamma_s $ is the
decay from the state $|S\rangle$ to the state $|0\rangle$. }
\label{decayfig}
\end{figure}

As it is shown in Fig. \ref{decayfig}, the state of the NV center $%
|A_2\rangle $ can decay to the state $|0\rangle$ with dark transition \cite%
{nature-478-497}, which gives neither heating nor cooling to the cantilever.
However, our cooling process will be stopped after the NV center is in the
state $|0\rangle$. To avoid suspension of the cooling process, we will apply
an additional pumping light on the transition from the state $|0\rangle$ to
the state $|E_y\rangle$. This pumping process can not only realize a
indirect transition from the state $|A_2\rangle$ to the state $|\pm1\rangle$%
, but also give an effective decay to this indirect transition.

The pumping four-level systems are present in the right-hand side in Fig. %
\ref{decayfig}, which start from the state $|0\rangle$ to the excited state $%
|E_{y}\rangle$, and then to the state $|\pm1\rangle$ or $|0\rangle$. The
decay processes for the final state $|\pm1\rangle$ and $|0\rangle$ are
corresponding to the effective decay process and dephasing process,
respectively. Then, the Hamiltonian describes the pumping process can be
given as ($\hbar =1$),

\begin{equation}
H_{p}=\omega _{e}\left\vert E_{y}\right\rangle \left\langle E_{y}\right\vert
+\omega _{+1}\left\vert +1\right\rangle \left\langle +1\right\vert +\omega
_{-1}\left\vert -1\right\rangle \left\langle -1\right\vert +\Omega
_{p}(\left\vert E_{y}\right\rangle \left\langle 0\right\vert +\left\vert
E_{y}\right\rangle \left\langle 0\right\vert )\cos (\omega _{p}t),
\label{APP1}
\end{equation}%
where the first three items describe the free energy for the NV center with $%
\omega _{e}$ and $\omega _{\pm 1}$ being the energy for states $\left\vert
E_{y}\right\rangle $ and $\left\vert \pm 1\right\rangle $, respectively; the
last item represents the transition between the states $\left\vert
E_{y}\right\rangle $ and $\left\vert 0\right\rangle $ driven by a pumping
field with a Rabi frequency $\Omega _{p}$ and frequency $\omega _{p}$.

According to the effective operator formalism for open quantum system \cite%
{pra-85-032111}, we can calculate the effective decays as follows.

In the rotating frame of the pumping field frequency $\omega _{p}$, the
above Hamiltonian can be rewritten as

\begin{equation}
\begin{array}{llllll}
H_{p} & = & H_{e}+H_{g}+V_{-}+V_{+}, &  &  &  \\
H_{e} & = & \Delta _{e}\left\vert E_{y}\right\rangle \left\langle
E_{y}\right\vert , & H_{g} & = & \omega _{\pm 1}\left\vert \pm
1\right\rangle \left\langle \pm 1\right\vert , \\
V_{-} & = & \frac{\Omega _{p}}{2}\left\vert 0\right\rangle \left\langle
E_{y}\right\vert , & V_{+} & = & \frac{\Omega _{p}}{2}\left\vert
E_{y}\right\rangle \left\langle 0\right\vert%
\end{array}
\label{APP2}
\end{equation}%
with $\Delta _{e}=\omega _{e}-\omega _{p}$.

Then the non-Hermitian Hamiltonian for the quantum jump formalism is
\begin{equation}
H_{NH}=H_{e}-\frac{i}{2}(\Gamma _{0}+\Gamma _{-1}+\Gamma
_{+1})(|E_{y}\rangle \langle E_{y}|).  \label{APP3}
\end{equation}%
And the corresponding effective Hamiltonian and Lindblad operators can be
given as:

\begin{equation}
\begin{array}{lll}
H_{eff} & = & -\frac{1}{2}V_{-}[H_{NV}^{-1}+(H_{NV}^{-1})^{\dag }]V_{+}+H_{g}
\\
& = & -\frac{\Delta _{e}\Omega _{p}^{2}}{4\Delta _{e}^{2}+(\Gamma
_{0}+\Gamma _{-1}+\Gamma _{+1})^{2}}\left\vert 0\right\rangle \left\langle
0\right\vert +\omega _{\pm 1}\left\vert \pm 1\right\rangle \left\langle \pm
1\right\vert ; \\
L_{op}^{k} & = & L_{k}H_{NV}^{-1}V_{+},L_{k}=\sqrt{\Gamma _{k}}\left\vert
k\right\rangle \left\langle E_{y}\right\vert ; \\
L_{op}^{0} & = & \sqrt{\Gamma _{0}}\left\vert 0\right\rangle \left\langle
E_{y}\right\vert \frac{\left\vert E_{y}\right\rangle \left\langle
E_{y}\right\vert }{\Delta _{e}-i\frac{1}{2}(\Gamma _{0}+\Gamma _{-1}+\Gamma
_{+1})}\frac{\Omega _{p}}{2}\left\vert E_{y}\right\rangle \left\langle
0\right\vert \\
& = & \sqrt{\Gamma _{0}}\frac{1}{\Delta _{e}-i\frac{1}{2}(\Gamma _{0}+\Gamma
_{-1}+\Gamma _{+1})}\frac{\Omega _{p}}{2}\left\vert 0\right\rangle
\left\langle 0\right\vert ; \\
L_{op^{\pm 1}} & = & \sqrt{\Gamma _{\pm 1}}\left\vert \pm 1\right\rangle
\left\langle E_{y}\right\vert \frac{\left\vert E_{y}\right\rangle
\left\langle E_{y}\right\vert }{\Delta _{e}-i\frac{1}{2}(\Gamma _{0}+\Gamma
_{-1}+\Gamma _{+1})}\frac{\Omega _{p}}{2}\left\vert E_{y}\right\rangle
\left\langle 0\right\vert \\
& = & \sqrt{\Gamma _{\pm 1}}\frac{1}{\Delta _{e}-i\frac{1}{2}(\Gamma
_{0}+\Gamma _{-1}+\Gamma _{+1})}\frac{\Omega _{p}}{2}\left\vert \pm
1\right\rangle \left\langle 0\right\vert .%
\end{array}
\label{APP4}
\end{equation}%
Here, $\Gamma _{0}$ ($\Gamma _{\pm 1}$) is the decay from the excited state $%
\left\vert E_{y}\right\rangle $ to the ground state $|0\rangle $ ($%
\left\vert \pm 1\right\rangle $).

As a result, the effective decay $L_{op}^{\pm 1}$ generates the ground state
$\left\vert \pm 1\right\rangle $ from $\left\vert 0\right\rangle $ can be
reduced by%
\begin{equation}
\begin{array}{lll}
\Gamma _{op}^{\pm 1} & = & |\left\langle \pm 1\right\vert \sqrt{\Gamma _{\pm
1}}\frac{1}{\Delta _{e}-i\frac{1}{2}(\Gamma _{0}+\Gamma _{-1}+\Gamma _{+1})}%
\frac{\Omega _{p}}{2}\left\vert \pm 1\right\rangle \left\langle 0\right\vert
\left\vert 0\right\rangle |^{2} \\
& = & \Gamma _{\pm 1}\frac{\Omega _{p}^{2}}{4\Delta _{e}^{2}+(\Gamma
_{0}+\Gamma _{-1}+\Gamma _{+1})^{2}}.%
\end{array}
\label{APP5}
\end{equation}%
and the dephasing of $\left\vert 0\right\rangle $ can be written as%
\begin{equation}
\begin{array}{lll}
\Gamma _{op}^{0} & = & |\left\langle 0\right\vert \sqrt{\Gamma _{0}}\frac{1}{%
\Delta _{e}-i\frac{1}{2}(\Gamma _{0}+\Gamma _{-1}+\Gamma _{+1})}\frac{\Omega
_{p}}{2}\left\vert 0\right\rangle \left\langle 0\right\vert \left\vert
0\right\rangle |^{2} \\
& = & \Gamma _{0}\frac{\Omega _{p}^{2}}{4\Delta _{e}^{2}+(\Gamma _{0}+\Gamma
_{-1}+\Gamma _{+1})^{2}}.%
\end{array}%
\end{equation}%
It is worth to point out that, $\Gamma _{op}^{+1}+\Gamma _{op}^{-1}+\Gamma
_{op}^{0}$ can't be larger than the decay for dark transition $\gamma
_{0}=\Gamma -\gamma _{+1}-\gamma _{-1}$ from the state $|A_{2}\rangle $ to
the state $|^{1}A_{1}\rangle $, and then to the state $\left\vert
0\right\rangle $, since both the effective decay and dephasing are original
from this dark transition process. Moreover, we can also define the real
decay from the excited state $\left\vert A_{2}\right\rangle $ to the ground
state $\left\vert \pm 1\right\rangle $ as $\gamma _{\pm }=\gamma _{\pm
1}+\Gamma _{op}^{\pm 1}$ with $\gamma _{\pm 1}$ being the direct decay from $%
\left\vert A_{2}\right\rangle $ to $\left\vert \pm 1\right\rangle $, and $%
\gamma _{+}+\gamma _{-}\leq \Gamma $. Considering the decay of dark
transition is small, for simplicity, we assume $\gamma _{+}+\gamma
_{-}\simeq \Gamma $ and use $\gamma _{\pm }$ to represent all the decays
from the excited state $|A_{2}\rangle $ to the ground state $|\pm 1\rangle $.

\begin{figure}[tbph]
\centering\includegraphics[width=6cm]{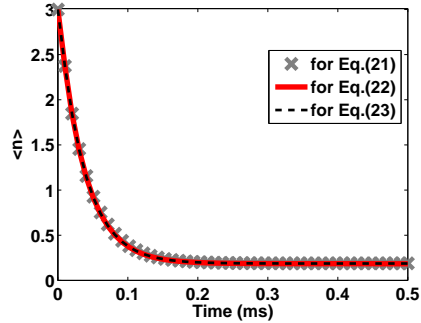} \centering
\caption{The average phonon number $\langle n\rangle $ as a function of the
time $T$. Assume the initial average phonon number is $\langle n\rangle =3$
and the NV center is in state $|-1\rangle $. The signs for crossing, the
black dash line and red solid line are simulated with Eqs.(\protect\ref{APP6}%
), (\protect\ref{APP7}) and (\protect\ref{APP8}), respectively. Here $%
\protect\omega _{m}=2\protect\pi \times 1$MHz, $\Omega _{0}=6\protect\omega %
_{m}$, $\Delta =10\protect\omega _{m}$, $\Gamma =15\protect\omega _{m}$, $%
\protect\gamma _{\pm 1}=\Gamma /2$, $\Gamma _{dark}=\Gamma /130$, $\Gamma
_{0}=\Gamma $, $\Gamma _{\pm 1}=\Gamma /150$, $\protect\gamma _{s}=\Gamma
/33 $, $\protect\gamma _{0}=0.1\Gamma $, $T=20$mK, $\protect\eta =0.115$, $%
\Delta _{p}=0 MHz$, and $\Omega _{p}=\Gamma $ \protect\cite%
{prb-79-041302,natphys-7-879,nature-466-730,njp-13-025025,pra-83-054306,arxiv1305.1701,nature-466-730,nature-478-497}%
.}
\label{nihe}
\end{figure}

To check the accuracy of the real decays, we simulate the average phonon
number $\left\langle n\right\rangle =Tr(b^{\dagger }b\rho )$ versus the time
$T$ in Fig. \ref{nihe} with and without the state $|0\rangle $. The
corresponding master equations for the density matrix $\rho $ without and
with the state $|0\rangle $, and the one for the real system are:

\begin{equation}
\begin{array}{lll}
\frac{d}{dt}\rho & = & -i[H^{\mathrm{rot}},\rho ]+\frac{\gamma _{m}}{2}%
[b\rho b^{\dagger }-\rho b^{\dagger }b-b^{\dagger }b\rho ] \\
& + & \sum\limits_{_{\pm }}\frac{\gamma _{\pm }}{2}[\left\vert \pm
1\right\rangle \left\langle A_{2}\right\vert \rho \left\vert
A_{2}\right\rangle \left\langle \pm 1\right\vert -\rho \left\vert
A_{2}\right\rangle \left\langle A_{2}\right\vert -\left\vert
A_{2}\right\rangle \left\langle A_{2}\right\vert \rho ],%
\end{array}
\label{APP6}
\end{equation}%
\begin{equation}
\begin{array}{lll}
\frac{d}{dt}\rho & = & -i[H^{\mathrm{rot}}-\omega _{0}\left\vert
0\right\rangle \left\langle 0\right\vert ,\rho ]+\frac{\gamma _{m}}{2}[b\rho
b^{\dagger }-\rho b^{\dagger }b-b^{\dagger }b\rho ] \\
& + & \sum\limits_{_{\pm }}\frac{\gamma _{\pm 1}}{2}[\left\vert \pm
1\right\rangle \left\langle A_{2}\right\vert \rho \left\vert
A_{2}\right\rangle \left\langle \pm 1\right\vert -\rho \left\vert
A_{2}\right\rangle \left\langle A_{2}\right\vert -\left\vert
A_{2}\right\rangle \left\langle A_{2}\right\vert \rho ] \\
& + & \frac{\gamma _{0}}{2}[\left\vert 0\right\rangle \left\langle
A_{2}\right\vert \rho \left\vert A_{2}\right\rangle \left\langle
0\right\vert -\rho \left\vert A_{2}\right\rangle \left\langle
A_{2}\right\vert -\left\vert A_{2}\right\rangle \left\langle
A_{2}\right\vert \rho ] \\
& + & \sum\limits_{_{\pm }}\frac{\Gamma _{op}^{\pm 1}}{2}[\left\vert \pm
1\right\rangle \left\langle 0\right\vert \rho \left\vert 0\right\rangle
\left\langle \pm 1\right\vert -\rho \left\vert 0\right\rangle \left\langle
0\right\vert -\left\vert 0\right\rangle \left\langle 0\right\vert \rho ],%
\end{array}
\label{APP7}
\end{equation}%
and%
\begin{equation}
\begin{array}{lll}
\frac{d}{dt}\rho & = & -i[H^{\mathrm{rot}}-\omega _{s}\left\vert
^{1}A_{1}\right\rangle \left\langle ^{1}A_{1}\right\vert +\Omega
_{p}(\left\vert E_{y}\right\rangle \left\langle 0\right\vert +\left\vert
0\right\rangle \left\langle E_{y}\right\vert ),\rho ] \\
& + & \frac{\gamma _{m}}{2}[b\rho b^{\dagger }-\rho b^{\dagger }b-b^{\dagger
}b\rho ] \\
& + & \sum\limits_{_{\pm }}\frac{\gamma _{\pm 1}}{2}[\left\vert \pm
1\right\rangle \left\langle A_{2}\right\vert \rho \left\vert
A_{2}\right\rangle \left\langle \pm 1\right\vert -\rho \left\vert
A_{2}\right\rangle \left\langle A_{2}\right\vert -\left\vert
A_{2}\right\rangle \left\langle A_{2}\right\vert \rho ] \\
& + & \frac{\Gamma _{dark}}{2}[\left\vert ^{1}A_{1}\right\rangle
\left\langle A_{2}\right\vert \rho \left\vert A_{2}\right\rangle
\left\langle ^{1}A_{1}\right\vert -\rho \left\vert A_{2}\right\rangle
\left\langle A_{2}\right\vert -\left\vert A_{2}\right\rangle \left\langle
A_{2}\right\vert \rho ] \\
& + & \frac{\gamma _{s}}{2}[\left\vert 0\right\rangle \left\langle
^{1}A_{1}\right\vert \rho \left\vert ^{1}A_{1}\right\rangle \left\langle
0\right\vert -\rho \left\vert ^{1}A_{1}\right\rangle \left\langle
^{1}A_{1}\right\vert -\left\vert ^{1}A_{1}\right\rangle \left\langle
^{1}A_{1}\right\vert \rho ], \\
& + & \frac{\Gamma _{0}}{2}[\left\vert 0\right\rangle \left\langle
E_{y}\right\vert \rho \left\vert E_{y}\right\rangle \left\langle
0\right\vert -\rho \left\vert E_{y}\right\rangle \left\langle
E_{y}\right\vert -\left\vert E_{y}\right\rangle \left\langle
E_{y}\right\vert \rho ], \\
& + & \sum\limits_{_{\pm }}\frac{\Gamma _{\pm 1}}{2}[\left\vert \pm
1\right\rangle \left\langle E_{y}\right\vert \rho \left\vert
E_{y}\right\rangle \left\langle \pm 1\right\vert -\rho \left\vert
E_{y}\right\rangle \left\langle E_{y}\right\vert -\left\vert
E_{y}\right\rangle \left\langle E_{y}\right\vert \rho ],%
\end{array}
\label{APP8}
\end{equation}%
respectively. The last term in Eq.(\ref{APP7}) describes the effective decay
process from the state $\left\vert 0\right\rangle $ to the state $\left\vert
\pm 1\right\rangle $, which is caused by the pumping process. Here, $\gamma
_{0}$ ($\gamma _{\pm 1}$, $\Gamma _{dark}$) is the decay from the excited
state $\left\vert A_{2}\right\rangle $ to the state $\left\vert
0\right\rangle $ ($\left\vert \pm 1\right\rangle $, $\left\vert
^{1}A_{1}\right\rangle $) , $\Gamma _{0}$ ($\Gamma _{\pm 1}$) is the decay
from the excited state $\left\vert E_{y}\right\rangle $ to the ground state $%
\left\vert 0\right\rangle $ ($\left\vert \pm 1\right\rangle $), $\gamma _{s}$
is the decay from the state $\left\vert ^{1}A_{1}\right\rangle $ to the
ground state $\left\vert 0\right\rangle $.

Compared with the simulation results, we find that the average phonon number
$\left\langle n\right\rangle =Tr(b^{\dagger }b\rho )$ versus the time $T$
with Eqs.(\ref{APP6}) (\ref{APP7}) and (\ref{APP8}) agree with each other
very well since the decay from $\left\vert A_{2}\right\rangle $ to the state
$\left\vert ^{1}A_{1}\right\rangle $ is very small, the three-level system
constitued by states $\left\vert A_{2}\right\rangle $ and $\left\vert \pm
1\right\rangle $ can be treated as a nearly closed three-level system. As a
result, the real decay of our model is valid.

\section*{Appendix B}

In what follows, we treat the coupling $V$ between the NV center and the
cantilever by perturbation theory and use the non-equilibrium
fluctuation-dissipation relation to derive the cooling and heating rates.

Defining operators $c=-ib$, $c^{\dagger }=ib^{\dagger }$ in Eq. (\ref{C1}),
we can get $\left\langle b^{\dagger }b\right\rangle =\left\langle c^{\dagger
}c\right\rangle $. After the position operator $X=x_{0}(c+c^{\dagger })$
introduced, we rewrite the interaction Hamiltonian (\ref{C2}) as
\begin{equation}
V=-\frac{i\eta }{x_{0}}X(\frac{\Omega _{0}}{\sqrt{2}}|A_{2}\rangle \langle
d|-h.c.),  \label{C4}
\end{equation}%
and the corresponding fluctuation spectrum is \cite{pra-46-2668}
\begin{equation}
S(\omega )=\frac{1}{2M\omega _{m}}\int_{0}^{\infty }\mathrm{dt}e^{i\omega
t}\langle F(t)F(0)\rangle _{ss},  \label{C5}
\end{equation}%
where the notation $\langle \cdots \rangle _{ss}$ stands for the average
value in atomic steady state $\rho _{ss}$ in the absence of cantilever and
the Heisenberg operator $F(t)$ takes the form of
\begin{equation}
F(t)=-\frac{d}{dX}V\mid _{X=0}=\frac{i\eta }{\sqrt{2}x_{0}}(\Omega
_{0}|A_{2}\rangle \langle d|-h.c.)=-\frac{\eta }{\sqrt{2}x_{0}}\Omega
_{0}\sigma _{y}^{A_{2},d}.  \label{C6}
\end{equation}%
Here $\sigma _{x}^{m,n}=|m\rangle \langle n|+|n\rangle \langle m|$ and $%
\sigma _{y}^{m,n}=-i(|m\rangle \langle n|-|n\rangle \langle m|)$ with $%
m,n=A_{2},\pm 1$.

The steady state for the NV center $\rho _{ss}$ can be obtained from the
Bloch equation for Hamiltonian (\ref{C1}) and can be given as \cite%
{pra-46-2668},
\begin{equation}
\begin{array}{ccl}
\frac{d\left\langle \rho ^{bb}\right\rangle }{dt} & = & -\frac{\sqrt{2}%
\Omega _{0}}{2}\left\langle \sigma _{y}^{A_{2},b}\right\rangle +\Gamma
_{b}(1-\left\langle \rho ^{bb}\right\rangle -\left\langle \rho
^{dd}\right\rangle ), \\
\frac{d\left\langle \rho ^{dd}\right\rangle }{dt} & = & \Gamma
_{d}(1-\left\langle \rho ^{bb}\right\rangle -\left\langle \rho
^{dd}\right\rangle ), \\
\frac{d\left\langle \sigma _{x}^{bd}\right\rangle }{dt} & = & -\frac{\sqrt{2}%
\Omega _{0}}{2}\left\langle \sigma _{y}^{A_{2},d}\right\rangle , \\
\frac{d\left\langle \sigma _{y}^{bd}\right\rangle }{dt} & = & \frac{\sqrt{2}%
\Omega _{0}}{2}\left\langle \sigma _{x}^{A_{2},d}\right\rangle , \\
\frac{d\left\langle \sigma _{x}^{A_{2},b}\right\rangle }{dt} & = & -\frac{%
\Gamma }{2}\left\langle \sigma _{x}^{A_{2},b}\right\rangle +\Delta
\left\langle \sigma _{y}^{A_{2},b}\right\rangle , \\
\frac{d\left\langle \sigma _{y}^{A_{2},b}\right\rangle }{dt} & = & -\frac{%
\Gamma }{2}\left\langle \sigma _{y}^{A_{2},b}\right\rangle -\sqrt{2}\Omega
_{0}(2\left\langle \rho ^{bb}\right\rangle +\left\langle \rho
^{dd}\right\rangle -1)-\Delta \left\langle \sigma
_{x}^{A_{2},b}\right\rangle , \\
\frac{d\left\langle \sigma _{x}^{A_{2},d}\right\rangle }{dt} & = & -\frac{%
\Gamma }{2}\left\langle \sigma _{x}^{A_{2},d}\right\rangle -\frac{\sqrt{2}%
\Omega _{0}}{2}\left\langle \sigma _{y}^{bd}\right\rangle +\Delta
\left\langle \sigma _{y}^{A_{2},d}\right\rangle , \\
\frac{d\left\langle \sigma _{y}^{A_{2},d}\right\rangle }{dt} & = & -\frac{%
\Gamma }{2}\left\langle \sigma _{y}^{A_{2},d}\right\rangle +\frac{\sqrt{2}%
\Omega _{0}}{2}\left\langle \sigma _{x}^{bd}\right\rangle -\Delta
\left\langle \sigma _{x}^{A_{2},d}\right\rangle ,%
\end{array}%
\end{equation}%
where $\gamma _{+}$, $\gamma _{-}$, $\Gamma _{d}=$ $\Gamma _{b}=(\gamma
_{+}+\gamma _{-})/2$ are the decay rates for the excited state $%
|A_{2}\rangle $ to the states $|+1\rangle $, $|-1\rangle $, $|d\rangle $,
and $|b\rangle $, respectively; $\Gamma =\gamma _{+}+\gamma _{-}$ is the
total decay rate; $\rho _{bb}=|b\rangle \langle b|$, $\rho _{dd}=|d\rangle
\langle d|$. The steady state for this Bloch equation is $\rho _{ss}=\rho
^{dd}$, which means the steady state for the NV center is in a dark state.

When the NV center is in its dark state, since only $\left\langle \sigma
_{y}^{A_{2},d}(t)\sigma _{y}^{A_{2},d}(0)\right\rangle _{ss}\neq 0$, the
fluctuation spectrum is reduced to
\begin{equation}
S(\omega )=\eta ^{2}(\frac{\Omega _{0}}{\sqrt{2}})^{2}\int_{0}^{\infty }%
\mathrm{dt}e^{i\omega t}\langle \sigma _{y}^{A_{2},d}(t)\sigma
_{y}^{A_{2},d}(0)\rangle _{ss}.  \label{C7}
\end{equation}%
According to the quantum regression theorem \cite{pra-46-2668}, the equation
of the correlation functions can be written as
\begin{equation}
\begin{array}{ccl}
\frac{d\left\langle \rho ^{bb}(t)\sigma _{y}^{A_{2},d}(0)\right\rangle _{ss}%
}{dt} & = & -\frac{\sqrt{2}\Omega _{0}}{2}\left\langle \sigma
_{y}^{A_{2},b}(t)\sigma _{y}^{A_{2},d}(0)\right\rangle _{ss}+\Gamma
_{b}(\left\langle \sigma _{y}^{A_{2},d}\right\rangle _{ss} \\
& - & \left\langle \rho ^{bb}(t)\sigma _{y}^{A_{2},d}(0)\right\rangle
_{ss}-\left\langle \rho ^{dd}(t)\sigma _{y}^{A_{2},d}(0)\right\rangle _{ss}),
\\
\frac{d\left\langle \rho ^{dd}(t)\sigma _{y}^{A_{2},d}(0)\right\rangle _{ss}%
}{dt} & = & \Gamma _{d}(\left\langle \sigma _{y}^{A_{2},d}\right\rangle
_{ss}-\left\langle \rho ^{bb}(t)\sigma _{y}^{A_{2},d}(0)\right\rangle
_{ss}-\left\langle \rho ^{dd}(t)\sigma _{y}^{A_{2},d}(0)\right\rangle _{ss}),
\\
\frac{d\left\langle \sigma _{x}^{b,d}(t)\sigma
_{y}^{A_{2},d}(0)\right\rangle _{ss}}{dt} & = & -\frac{\sqrt{2}\Omega _{0}}{2%
}\left\langle \sigma _{y}^{A_{2},d}(t)\sigma _{y}^{A_{2},d}(0)\right\rangle
_{ss}, \\
\frac{d\left\langle \sigma _{y}^{b,d}(t)\sigma
_{y}^{A_{2},d}(0)\right\rangle _{ss}}{dt} & = & \frac{\sqrt{2}\Omega _{0}}{2}%
\left\langle \sigma _{x}^{A_{2},d}(t)\sigma _{y}^{A_{2},d}(0)\right\rangle
_{ss}, \\
\frac{d\left\langle \sigma _{x}^{A_{2},b}(t)\sigma
_{y}^{A_{2},d}(0)\right\rangle _{ss}}{dt} & = & -\frac{\Gamma }{2}%
\left\langle \sigma _{x}^{A_{2},b}(t)\sigma _{y}^{A_{2},d}(0)\right\rangle
_{ss}+\Delta \left\langle \sigma _{y}^{A_{2},b}(t)\sigma
_{y}^{A_{2},d}(0)\right\rangle _{ss}, \\
\frac{d\left\langle \sigma _{y}^{A_{2},b}(t)\sigma
_{y}^{A_{2},d}(0)\right\rangle _{ss}}{dt} & = & -\frac{\Gamma }{2}%
\left\langle \sigma _{y}^{A_{2},b}(t)\sigma _{y}^{A_{2},d}(0)\right\rangle
_{ss}+\sqrt{2}\Omega _{0}(2\left\langle \rho ^{bb}(t)\sigma
_{y}^{A_{2},d}(0)\right\rangle _{ss} \\
& + & \left\langle \rho ^{dd}(t)\sigma _{y}^{A_{2},d}(0)\right\rangle
_{ss}-\left\langle \sigma _{y}^{A_{2},d}\right\rangle _{ss})-\Delta
\left\langle \sigma _{x}^{A_{2},b}(t)\sigma _{y}^{A_{2},d}(0)\right\rangle
_{ss}, \\
\frac{d\left\langle \sigma _{x}^{A_{2},d}(t)\sigma
_{y}^{A_{2},d}(0)\right\rangle _{ss}}{dt} & = & -\frac{\Gamma }{2}%
\left\langle \sigma _{x}^{A_{2},d}(t)\sigma _{y}^{A_{2},d}(0)\right\rangle
_{ss}-\frac{\sqrt{2}\Omega _{0}}{2}\left\langle \sigma _{y}^{b,d}(t)\sigma
_{y}^{A_{2},d}(0)\right\rangle _{ss} \\
& + & \Delta \left\langle \sigma _{y}^{A_{2},d}(t)\sigma
_{y}^{A_{2},d}(0)\right\rangle _{ss}, \\
\frac{d\left\langle \sigma _{y}^{A_{2},d}(t)\sigma
_{y}^{A_{2},d}(0)\right\rangle _{ss}}{dt} & = & -\frac{\Gamma }{2}%
\left\langle \sigma _{y}^{A_{2},d}(t)\sigma _{y}^{A_{2},d}(0)\right\rangle
_{ss}+\frac{\sqrt{2}\Omega _{0}}{2}\left\langle \sigma _{x}^{b,d}(t)\sigma
_{y}^{A_{2},d}(0)\right\rangle _{ss} \\
& - & \Delta \left\langle \sigma _{x}^{A_{2},d}(t)\sigma
_{y}^{A_{2},d}(0)\right\rangle _{ss}.%
\end{array}
\label{C8}
\end{equation}

Define the transformation
\begin{equation}
f(t)\rightleftharpoons F(\nu )=\int_{0}^{\infty }\mathrm{dt}e^{i\nu t}f(t),
\end{equation}%
then
\begin{equation}
\frac{f(t)}{dt}\rightleftharpoons -f(0)-i\nu F(\nu ).
\end{equation}%
After the equations above solved, we can obtain
\begin{equation}
\int_{0}^{\infty }\mathrm{dt}e^{i\omega t}\left\langle \sigma
_{y}^{A_{2},d}(t)\sigma _{y}^{A_{2},d}(0)\right\rangle _{ss}=\frac{2i\omega
}{i\Gamma \omega +2\Delta \omega +2\omega ^{2}-\Omega _{0}^{2}}  \label{C09}
\end{equation}%
and the corresponding heating (cooling) coefficient $A_{+}$ ($A_{-}$) as
\begin{eqnarray}
A_{\pm } &=&2\mathrm{Re}\{S(\mp \omega _{m})\}  \nonumber \\
&=&2\Gamma \eta ^{2}\Omega _{0}^{2}\frac{\omega _{m}^{2}}{\Gamma ^{2}\omega
_{m}^{2}+4[\frac{\Omega _{0}^{2}}{2}\pm \Delta \omega _{m}-\omega
_{m}^{2}]^{2}}.  \label{C2001}
\end{eqnarray}

\section*{Acknowledgments}

J.Q.Z would like to thank Yi Zhang, Zhang-Qi Yin, Chuan-Jia Shan, Keyu Xia
and Nan Zhao for valuable discussions. The work is supported by the National
Fundamental Research Program of China (Grant Nos. 2012CB922102,
2012CB922104, and 2009CB929604), the National Natural Science Foundation of
China (Grant Nos. 10974225, 60978009, 11174027, 11174370, 11274036,
11304366, 11304174, {11322542 }and 61205108), and the China Postdoctoral
Science Foundation (Grant No. 2013M531771).

\begin{thebibliography}{99}
\bibitem{V.B.Braginsky} V. B. Braginsky and A. B. Manukin, {}``Measurements
of Weak Forces in Physics Experiments,\textquotedblright{}\ D. H. Douglass
eds. (Chicago University Press, Chicago, 1977).

\bibitem{prl-97-237201} L. F. Wei, Y. X. Liu, C. P. Sun, and F. Nori,
{}``Probing tiny nanomechanical resonator: classical or quantum mechanical?,%
\textquotedblright{}\ Phys. Rev. Lett. \textbf{97}, 237201 (2006).

\bibitem{Physics-2-40} F. Marquardt and S. M. Girvin, {}``Optomechanics,%
\textquotedblright{}\ Physics \textbf{2}, 40 (2009).

\bibitem{pra-84-024301} H. T. Tan and G. X. Li, {}``Multicolor quadripartite
entanglement from an optomechanical cavity,\textquotedblright{} Phys. Rev. A
\textbf{84},\ 024301 (2011).

\bibitem{prl-101-200503} M. J. Hartmann and M. B. Plenio, {}``Steady State
Entanglement in the Mechanical Vibrations of Two Dielectric Membranes,%
\textquotedblright{}\ Phys. Rev. Lett. \textbf{101},\ 200503 (2008).

\bibitem{prl-108-120801} S. Forstner, S. Prams, J. Knittel, E. D. van
Ooijen, J. D. Swaim, G. I. Harris, A. Szorkovszky, W. P. Bowen, and H.
Rubinsztein-Dunlop, {}``Cavity Optomechanical Magnetometer,%
\textquotedblright{}\ Phys. Rev. Lett. \textbf{108},\ 120801 (2012).

\bibitem{pra-86-053806} J. Q. Zhang, Y. Li, M. Feng, and Y. Xu,
{}``Precision measurement of electrical charge with optomechanically induced
transparency,\textquotedblright{}\ Phys. Rev. A \textbf{86},\ 053806 (2012).

\bibitem{prl-105-220501} K. Stannigel, P. Rabl, A. S. Sorensen, P. Zoller,
and M. D. Lukin, {}``Optomechanical Transducers for Long-Distance Quantum
Communication,\textquotedblright{}\ Phys. Rev. Lett. \textbf{105}, 220501
(2010).

\bibitem{Nat.nanotech-3-501} L. Tetard, A. Passian, K. T. Venmar, R. M.
Lynch, B. H. Voy, G. Shekhawat, V. P. Dravid, and T. Thundat, {}``Imaging
nanoparticles in cells by nanomechanical holography,\textquotedblright{}\
Nat. Nanotechnol. \textbf{3},\ 501-505 (2008).

\bibitem{prl-92-075507} I. Wilson-Rae, P. Zoller, and A. Imamoglu, {}``Laser
Cooling of a Nanomechanical Resonator Mode to its Quantum Ground State,%
\textquotedblright{}\ Phys. Rev. Lett. \textbf{92}, 075507 (2004).

\bibitem{Nature-475-359} J. D. Teufel, T. Donner, D. Li, J. W. Harlow, M. S.
Allman, K. Cicak, A. J. Sirois, J. D. Whittake, K. W. Lehnert, and R. W.
Simmonds, {}``Sideband cooling of micromechanical motion to the quantum
ground state,\textquotedblright{}\ Nature (London) \textbf{475}, 359-363 (2011).

\bibitem{PRB-78-134301} Y. Li, Y. D. Wang, F. Xue, and C. Bruder,
{}``Quantum theory of transmission line resonator-assisted cooling of a
micromechanical resonator,\textquotedblright{}\ Phys. Rev. B \textbf{78},\
134301 (2008).

\bibitem{Nature-432-200} J. D. Thompson, B. M. Zwickl, A. M. Jayich, F.
Marquardt, S. M. Girvin, and J. G. E. Harris, {}``Strong dispersive coupling
of a high-finesse cavity to a micromechanical membrane,\textquotedblright{}\
Nature (London) \textbf{452},\ 72-75 (2008).

\bibitem{prl-99-093901} I. Wilson-Rae, N. Nooshi, W. Zwerger, and T. J.
Kippenberg, {}``Theory of Ground State Cooling of a Mechanical Oscillator
Using Dynamical Backaction,\textquotedblright{}\ Phys. Rev. Lett. \textbf{99}%
,\ 093901 (2007).

\bibitem{prl-99-093902} F. Marquardt, J. P. Chen, A. A. Clerk, and S. M.
Girvin, {}``Quantum Theory of Cavity-Assisted Sideband Cooling of Mechanical
Motion,\textquotedblright{}\ Phys. Rev. Lett. \textbf{99}, 093902 (2007).

\bibitem{prb-76-205302} F. Xue, Y. D. Wang, Y. X. Liu, and F. Nori,
{}``Cooling a Micro-mechanical Beam by Coupling it to a Transmission Line,%
\textquotedblright{}\ Phys. Rev. B \textbf{76},\ 205302 (2007).

\bibitem{JPCM-25-142201} J. -Q. Zhang, Y. Li, and M. Feng, {}``Cooling a
charged mechanical resonator with time-dependent bias gate voltages,%
\textquotedblright{}\ J. Phys.: Condens. Matter \textbf{25}, 142201 (2013).

\bibitem{prl-108-120602} A. Mari and J. Eisert, {}``Very Hot Thermal Light
Can Significantly Cool Quantum Systems,\textquotedblright{} Phys. Rev. Lett.
\textbf{108},\ 120602 (2012).

\bibitem{pra-83-043804} Y. Li, L. A. Wu, and Z. D. Wang, {}``Fast
ground-state cooling of mechanical resonators with time-dependent optical
cavities.,\textquotedblright{}\ Phys. Rev. A \textbf{83},\ 043804 (2011).

\bibitem{pra-85-025804} Z. J. Deng, Y. Li, and C. W. Wu, {}``Performance of
a cooling method by quadratic coupling at high temperatures,%
\textquotedblright{}\ Phys. Rev. A \textbf{85,}\ 025804 (2012),

\bibitem{PRB-84-094502} Y. Li, L. A. Wu, Y. D. Wang, and L. P. Yang,
{}``Nondeterministic ultrafast ground-state cooling of a mechanical
resonator,\textquotedblright{}\ Phys. Rev. B \textbf{84},\ 094502 (2011).

\bibitem{Dissipative-Cooling-PRL2013} Y.-C. Liu, Y.-F. Xiao, X. Luan, C. W.
Wong, Phys. Rev. Lett. \textbf{110}, 153606 (2013).

\bibitem{Nature-464-697} A. D. O'Connell, M. Hofheinz, M. Ansmann, R. C.
Bialczak, M. Lenander, E. Lucero, M. Neeley, D. Sank, H. Wang, M. Weides, J.
Wenner, J. M. Martinis, and A. N. Cleland, {}``Quantum ground state and
single-phonon control of a mechanical resonator,\textquotedblright{}\ Nature
(London) \textbf{464},\ 697-703 (2010).

\bibitem{Nature-478-89} J. Chan, T. P. M. Alegre, A. H. Safavi-Naeini, J. T.
Hill, A. Krause, S. Grblacher, M. Aspelmeyer, and O. Painter,
{}\textquotedblleft Laser cooling of a nanomechanical oscillator into its
quantum ground state,\textquotedblright {}\ Nature (London) \textbf{478},
89-92 (2011).

\bibitem{prl-85-5547} C. F. Roos, D. Leibfried, A.Mundt, F. Schmidt-Kaler,
J. Eschner, and R. Blatt, {}``Experimental Demonstration of Ground State
Laser Cooling with Electromagnetically Induced Transparency,%
\textquotedblright{}\ Phys. Rev. Lett. \textbf{85},\ 5547-5550 (2000).

\bibitem{prl-85-4458} G. Morigi, J. Eschner, and C. H. Keitel,
{}\textquotedblleft Ground State Laser Cooling with Electromagnetically
Induced Transparency,\textquotedblright {}\ Phys. Rev. Lett. \textbf{85},\
4458-4461 (2000); G. Morigi, {}\textquotedblleft Cooling atomic motion with
quantum interference,\textquotedblright {}\ Phys. Rev. A \textbf{67},\
033502 (2003).

\bibitem{njp-9-279} A. Retzker and M. B. Plenio, {}``Fast cooling of trapped
ions using the dynamical Stark shift,\textquotedblright{} New J. Phys.
\textbf{9},\ 279 (2007).

\bibitem{prl-103-227203} K. Xia and J. Evers, {}``Ground State Cooling of a
Nanomechanical Resonator in the Nonresolved Regime via Quantum Interference,%
\textquotedblright{}\ Phys. Rev. Lett. \textbf{103}, 227203 (2009).

\bibitem{natnano} N. M. Nusran, M. Ummal Momeen, and M. V. Gurudev Dutt1
{}``High-dynamic-range magnetometry with a single electronic spin in diamond,%
\textquotedblright{}\ Nat. Nanotech. \textbf{7}, 109-113, (2012)

\bibitem{prb-79-041302} P. Rabl, P. Cappellaro, M. V. Gurudev Dutt, L.
Jiang, J. R. Maze, and M. D. Lukin, {}\textquotedblleft Strong magnetic
coupling between an electronic spin qubit and a mechanical
resonator,\textquotedblright {}\ Phys. Rev. B \textbf{79},\ 041302 (2009).

\bibitem{natphys-7-879} O. Arcizet, V. Jacques, A. Siria, P. Poncharal, P.
Vincent, and S. Seidelin, {}\textquotedblleft A single nitrogen-vacancy
defect coupled to a nanomechanical oscillator,\textquotedblright {}\ Nature
Phys. \textbf{7},\ 879-883 (2011).


\bibitem{rmp-77-633} M. Fleischhauer, A. Imamoglu, and J. P. Marangos,
{}\textquotedblleft Electromagnetically induced transparency: Optics in
coherent media, \textquotedblright {}\ Rev. Mod. Phys. \textbf{77},\ 633-673
(2005).

\bibitem{prl-104-043003} J. Cerrillo, A. Retzker, and M. B. Plenio,
\textquotedblleft Fast and Robust Laser Cooling of Trapped
Systems,\textquotedblright {}\ Phys. Rev. Lett. \textbf{104},\ 043003 (2009);


\bibitem{nature-466-730} E. Togan, Y. Chu, A. S. Trifonov, L. Jiang, J.
Maze, L. Childress, M. V. G. Dutt, A. S. S{\o }rensen, P. R. Hemmer, A. S.
Zibrov, and M. D. Lukin, {}\textquotedblleft Quantum entanglement between an
optical photon and a solid-state spin qubit, \textquotedblright {}\ Nature
\textbf{466},\ 730-734 (2010).

\bibitem{nature-478-497} E. Togan, Y. Chu, A. Imamoglu and, M. D. Lukin,
{}``Laser cooling and real-time measurement of the nuclear spin environment
of a solid-state qubit, \textquotedblright{}\ Nature \textbf{478},\ 497-501
(2011).

\bibitem{njp-13-025025} J. R. Maze, A. Gali, E. Togan, Y. Chu, A. Trifonov, E.
Kaxiras, and M. D. Lukin, {}``Properties of nitrogen-vacancy centers in
diamond: the group theoretic approach,\textquotedblright{}\ New J. Phys.
\textbf{13},\ 025025 (2011).

\bibitem{nature-304-74} M. D. LaHaye, O. Buu, B. Camarota, and K. Schwab,
{}``Quantum entanglement between an optical photon and a solid-state spin
qubit,\textquotedblright{}\ Nature (London) \textbf{466}, 730-734 (2010).

\bibitem{pra-83-054306} Q. Chen, W. L. Yang, M. Feng, and J. F. Du,
{}``Entangling separate nitrogen-vacancy centers in a scalable fashion via
coupling to microtoroidal resonators,\textquotedblright{}\ Phys. Rev. A
\textbf{83},\ 054305 (2011).

\bibitem{prl-87-257904} F. Mintert and C. Wunderlich, {}`` Ion-trap quantum
logic using long-wavelength radiation,\textquotedblright{}\ Phys. Rev. Lett.
\textbf{87}, 257904 (2001).

\bibitem{nphys-6-602} P. Rabl, S. J. Kolkowitz, F. H. L. Koppens, J. G. E.
Harris, P. Zoller, and M. D. Lukin, {}``A quantum spin transducer based on
nanoelectromechanical resonator arrays,\textquotedblright{}\ Nature Phys.
\textbf{6},\ 602-608 (2010).

\bibitem{Gardiner} S. A. Gardiner, {}``Dissertation: Quantum
Measurement, Quantum Chaos, and Bose-Einstein
Condensates,\textquotedblright\ (2000).

\bibitem{pra-72-043823} P. Rabl, V. Steixner, and P. Zoller,
\textquotedblleft Quantum-limited velocity readout and quantum feedback
cooling of a trapped ion via electromagnetically induced
transparency,\textquotedblright {}\ Phys. Rev. A. \textbf{72},\ 043823 (2005).

\bibitem{arxiv1305.1701} Z.-Q. Yin, T.-Z Li, X. Zhang, and L. -M. Duan,
{}``Large quantum superpositions of a levitated nanodiamond through
spin-optomechanical coupling,\textquotedblright{}\ Phys. Rev. A \textbf{88},
033614 (2013).


\bibitem{prl-102-057403} V. Jacques, P. Neumann, J. Beck, M. Markham, D.
Twitchen, J. Meijer, F. Kaiser, G. Balasubramanian, F. Jelezko, and J.
Wrachtrup, \textquotedblleft Dynamic Polarization of Single Nuclear Spins by
Optical Pumping of Nitrogen-Vacancy Color Centers in Diamond at Room
Temperature,\textquotedblright {}\ Phys. Rev. Lett. \textbf{102}, 057403
(2009).

\bibitem{NL-12-2083} T. Ishikawa, K.-M. C. Fu, C. Santori, V. M. Acosta, R.
G. Beausoleil, H. Watanabe, S. Shikata, and K. M. Itoh, \textquotedblleft
Optical and spin coherence properties of nitrogen-vacancy centers placed in
a 100 nm thick isotopically purified diamond layer,\textquotedblright {}\,
Nano. Lett. \textbf{12}, 2083-2087 (2012).


\bibitem{pra-85-032111} F. Reiter, and A. S. Sorensen, \textquotedblleft
Effective operator formalism for open quantum systems,\textquotedblright {}\
Phys. Rev. A \textbf{85}, 032111 (2012).


\bibitem{pra-46-2668} J. I. Cirac, R. Blatt, and P. Zoller,
{}\textquotedblleft Laser cooling of trapped ions in a standing
wave,\textquotedblright {}\ Phys. Rev. A \textbf{46},\ 2668-2681 (1992).
\end{thebibliography}
\end{document}